# High-contrast dark resonance on the $D_2$ - line of $^{87}$Rb in a vapor cell with different directions of the pump - probe waves

G.Kazakov[1*], I.Mazets[2], Yu.Rozhdestvensky[3],
G.Mileti[4], J.Delporte[5], and B.Matisov[1]

[1]*St. Petersburg State Polytechnic University,
Polytekhnicheskaya 29, St. Petersburg 195251, Russia*

[2]*A.F. Ioffe Physics-Technical Institute,
Polytekhnicheskaya 26, St. Petersburg 194021, Russia*

[3]*Institute for Laser Physics,
Birzhevaya liniya 12, St. Petersburg 199034, Russia*

[4]*Observatoire Cantonal Neuchatel,
rue de l'Observatoire 58, CH-2000, Neuchatel, Switzerland*

[5]*Centre National d'Etudes Spatiales (CNES)
avenue Belin 18, Toulouse 31401, France*

**Abstract**

We propose a novel method enabling to create a high-contrast dark resonance in the $^{87}$Rb vapor $D_2$-line. The method is based on an optical pumping of atoms into the working states by a two-frequency, linearly-polarized laser radiation propagating perpendicularly to the probe field. This new scheme is compared to the traditional scheme involving the circularly-polarized probe beam only, and significant improvement of the dark resonance parameters is found. Qualitative considerations are confirmed by numerical calculations.

---

[*] E-mail: kazakov@quark.stu.neva.ru

High-precision quantum frequency standards (atomic clocks) become now more and more indispensable in various scientific and engineering applications. A new boost of efforts to create a quantum frequency standard based on the coherent population trapping (CPT) effect has recently begun.

The CPT effect of atoms interacting with a resonant electromagnetic field is well known (see reviews [1,2] and references therein) and widely used in various fields of atomic and laser physics. Efforts to design and build CPT-based atomic clocks were undertaken by several scientific groups [3-8]. The main advantage of quantum frequency standard based on the CPT effect is that the corresponding RF resonance (so-called dark resonance) is excited by all-optical methods. However, practical use of the CPT phenomenon requires thorough optimization of the two-photon resonance parameters (width, amplitude, contrast). Here we propose a new method enabling to improve significantly the characteristics of the signal, due to very efficient accumulation of atomic population in the two working sublevels in a particular optical pumping scheme.

In CPT based clocks, the Zeeman sublevels $|1\rangle = |F=1, m=0\rangle$ and $|2\rangle = |F=2, m=0\rangle$ of the two ground-states hyperfine components of an alkali metal atom ($^{87}$Rb) are the working energy levels. The coherence between these states is created by a two-photon Raman transition induced by a circularly-polarized laser field containing two components with frequencies $\omega_1$ and $\omega_2$

$$\vec{E}(z,t) = \frac{\vec{e}_{+1}}{2}(E_1 \exp[i(k_1 z - \omega_1 t)] + E_2 \exp[i(k_2 z - \omega_2 t)]) + c.c.$$

in the presence of a level-splitting static magnetic field oriented along the $z$ axis. Here $\vec{e}_{+1} = -(\vec{e}_x + i\vec{e}_y)/\sqrt{2}$ is the unit vector of right-handed circular polarization and $E_{1,2}$ are the amplitudes of corresponding frequency components. Let us introduce the two-photon detuning $\Omega$ as

$$\Omega = (\Delta_{hfs} - (\omega_1 - \omega_2))/2,$$

where $\Delta_{hfs}$ is the ground state hyperfine splitting. When $\Omega$ is scanned across zero, a narrow dip (dark resonance) is observed in the absorption spectrum. The dark resonance width $\Gamma_s$ is determined by the relaxation rate of the ground state atomic coherence and the laser field parameters [1,2]. However, in the case of circularly-polarized fields there is always a trapping state (so-called "pocket") for an alkaline atom. E. g., for a $\sigma^+$-polarized laser radiation this is the sublevel with the maximum projection on the axis $z$ of the total angular momentum. Atoms are accumulated in the pocket and do not contribute anymore to the signal formation on the working transition. It leads to a drastic decrease of the signal amplitude and contrast.

To solve this problem, a scheme involving counter-propagating light waves with orthogonal polarizations has been proposed [5]. In that case, the pocket is absent. However, a significant periodic spatial variation of the resonance amplitude emerges. The period $\pi/(k_1 - k_2)$, where $k_j = 2\pi/\lambda_j$ is the wave number of the laser field of frequency $\omega_j$, is due to the difference of the wavelengths $\lambda_j$ of the two frequency components of the laser radiation. For example, this period is equal to approximately 2 cm for $^{87}$Rb atoms. Therefore spatial averaging over the cell decreases the dark resonance amplitude for the cell length $L \geq \pi/(k_1 - k_2)$ [6]. On the other hand, small cells of the length $L \ll \pi/(k_1 - k_2)$ are characterized by large values of the atomic relaxation at cell walls. To decrease the adverse influence of the wall relaxation, one needs to apply high buffer gas pressure (150 Torr in a cell of 12 mm$^3$ volume [7]). However, there are also inelastic collisions of the alkali atoms with the buffer gas resulting in decrease of the amplitude and significant broadening of the dark resonance. Thus the scheme proposed in [5] yielded practically only a small (1.4-fold) increase of the amplitude [7] compared to the usual co-propagating light waves configuration.

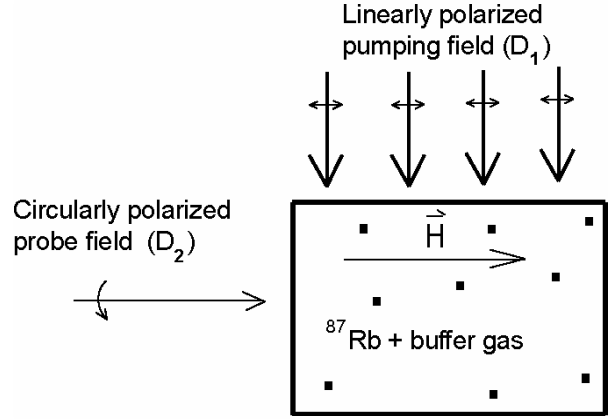

Figure 1: The proposed Rb Gas Cell Frequency Standard for the novel transversal pumping scheme

In order to preserve atoms from leaving the working sublevels and thus enhance the amplitude and contrast, we propose a new method, based on the use of a two-frequency optical pumping, as shown in Fig.1. (pumping scheme with a similar spatial configuration was considered in another context [9])

Let us suppose that $^{87}$Rb atoms in a cell interact with the linearly-polarized laser field (the pumping field)

$$\vec{E}_{pump} = \frac{\vec{e}_z}{2}\left(E_3 \exp[i(k_3 x - \omega_3 t)] + E_4 \exp[i(k_4 x - \omega_4 t)]\right) + c.c.$$

One of the components of the pumping field is tuned in resonance with the $|F=2\rangle \to |F'=2\rangle$ transition and the second component is tuned in resonance with the $|F=1\rangle \to |F'=1\rangle$ transition, both of these transitions belonging to the D$_1$-line (see Fig.2) In this case, atoms are not pumped out from the working levels $|1\rangle = |F=1, m=0\rangle$ and $|2\rangle = |F=2, m=0\rangle$, since the corresponding Clebsch-Gordan coefficients and, hence, transition dipole matrix elements are equal to zero [10]. As a result, after a few optical pumping cycles (coherent excitation followed by spontaneous relaxation of the optically excited state) all the atoms will be accumulated in the working sublevels $|1\rangle$ and $|2\rangle$. Note that the two frequency components of the pumping field don't need to be correlated.

Suppose now that we switch on the weak (compared to the pumping field) probe field

$$\vec{E}_{probe} = \frac{\vec{e}_{+1}}{2}\left(E_1 \exp[i(k_1 z - \omega_1 t)] + E_2 \exp[i(k_2 z - \omega_2 t)]\right) + c.c.,$$

which is circularly polarized and consists of two correlated frequency components tuned in resonance with the transitions from the ground state components to the excited states with $J' = 3/2$ (see Fig. 2). Since all the atoms are accumulated in the working sublevels, the contrast and amplitude of the probe beam absorption signal are greatly enhanced.

We have performed numerical calculations of the absorption in a $^{87}$Rb vapor cell at room temperature. The density matrix approach is used. We take into account the real hyperfine (hf) and Zeeman structure of the states involved as well as the exact values of the probabilities of the optically-induced one-photon transitions and spontaneous relaxation of the optically excited states. Effects of the thermal motion of atoms (Doppler broadening) also have been taken into account.

For $^{87}$Rb at 300 K the Doppler widths of the $D_1$ and $D_2$ lines are about 400 MHz, what is more then the hf splitting of the excited states with $J' = 3/2$, but less then the hf splitting of the excited states with $J' = 1/2$ (about 800 MHz).

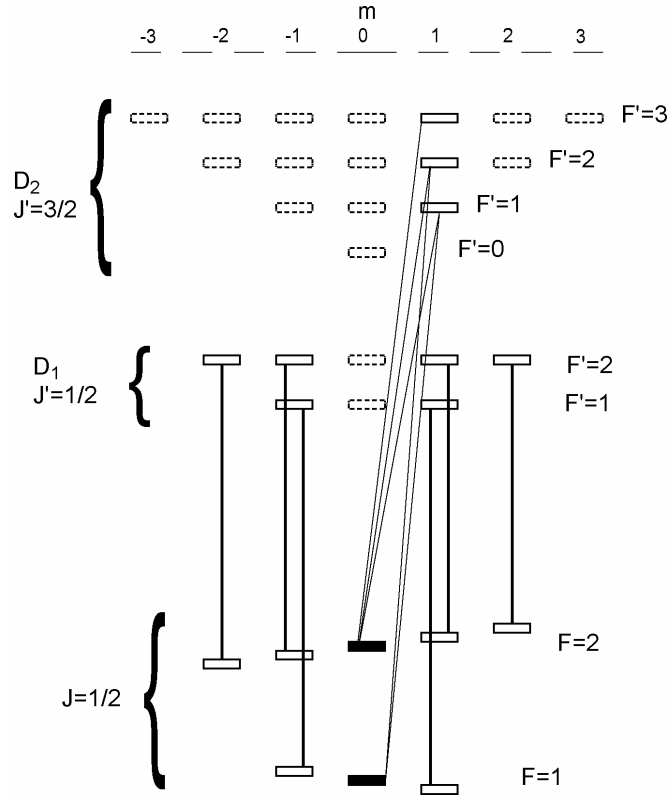

Figure2: The scheme of optically-induced transitions between the energy levels of $^{87}$Rb. Atoms are accumulated in the two working levels (black rectangles) due to optical pumping via the transitions indicated by bold vertical lines and spontaneous relaxation. The transitions excited by the probe field are shown by thin inclined lines.

We assume that the collision quenching of the excited state is negligible. Therefore optical relaxation (i.e. population relaxation from excited state to the different sublevels of the ground state) is provided by spontaneous transitions with the rate $\gamma = \gamma_{sp} \approx 3.6 \cdot 10^7 s^{-1}$ both for the $D_1$- and $D_2$-line. Relaxation of coherences between ground and excited states is assumed to be equal to $\gamma_{sp}/2$. Also we suppose that the vapor cell contains a buffer gas that provides Dicke narrowing of the hf transition [8].

Relaxation rate $\Gamma$ of coherences between the different hyperfine- and Zeeman sublevels of the ground state was supposed to be equal to $\Gamma \approx 200\ s^{-1}$. The depolarization rate in ground state is supposed to be equal to $\Gamma$.

Let us denote the total population in excited states (for the $\sigma^+$ excitation scheme), or the total population on all the sublevels of the $J' = 3/2$ excited state (for the scheme with transversal pumping) as $\rho_e$. We find $\rho_e$ from the density matrix equations. Then the absorbed power $\delta U_{probe}$ of probe field is equal to:

$$\delta U_{probe} = \gamma \cdot \hbar \omega \cdot N \cdot \rho_e ,$$

where $\hbar \omega$ is the optical transition energy, $N$ is a total number of active atoms. The "absorbed photodetector current" is equal to the difference between the photodetector current without ($I_0$) and with ($I$) the gas cell (see Fig.3):

$$I_0 - I = \delta U_{probe} \cdot \frac{e}{\hbar \omega} \cdot \kappa = \gamma \cdot e \cdot N \cdot \rho_e \cdot \kappa .$$

Here $e$ is elementary charge, $\kappa \approx 1$ — effective quantum efficiency. In our calculations we suppose that the vapor cell contains $N = 10^{11}$ Rb atoms, that is typical for a cell of volume ~10 cm$^3$ at 300 K [11].

The *amplitude A* of the dark resonance is defined as the difference between the current $I_r$ on the two-photon resonance condition and the same current $I_{nr}$ far from this resonance. *The contrast C* is the ratio of the amplitude A to the absorbed current $I_0 - I_{nr}$ far from resonance.

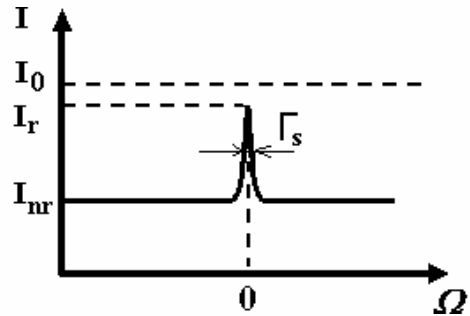

Figure 3: Qualitative picture showing the dark resonance line shape. $\Gamma_S$ is a width of dark resonance

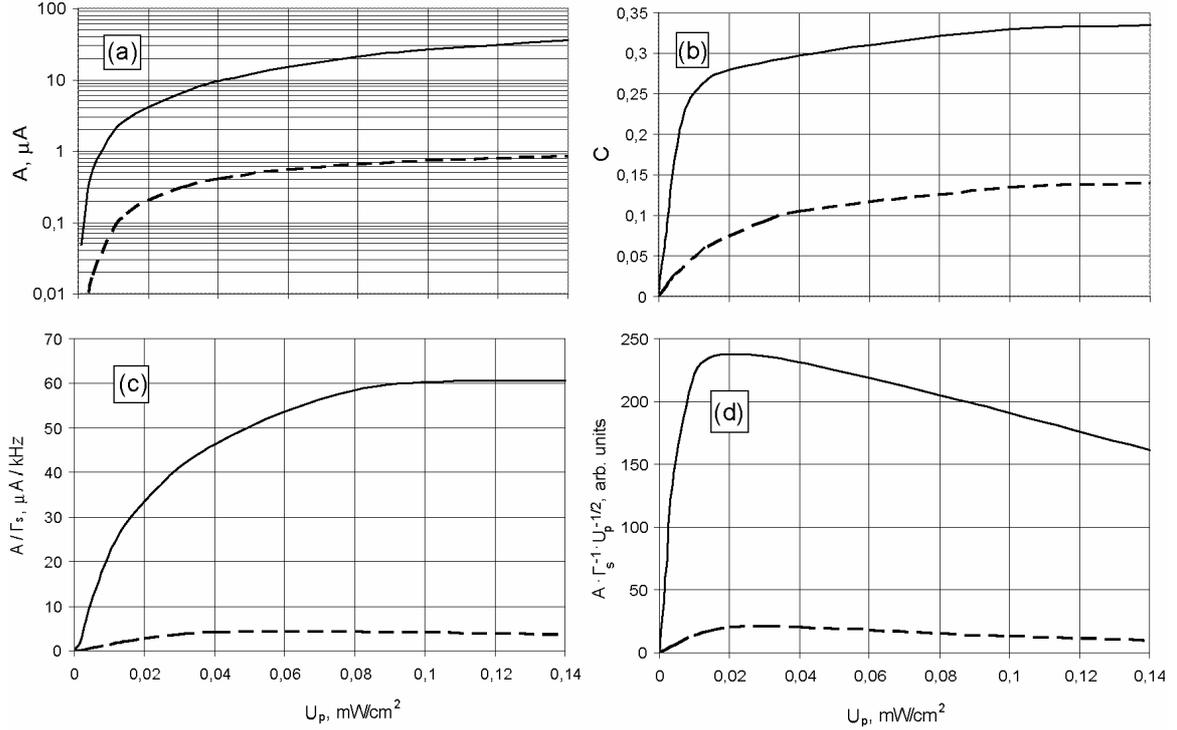

Figure 4: (a) Amplitude, (b) contrast, (c) amplitude-to-width ratio, (d) amplitude-to to-width ratio, divided by the square root of the probe field intensity of the dark resonance versus probe field intensity $U_P$. Solid line is for the novel scheme involving perpendicular pumping beam with the intensity $U_0 = 10\, mW/cm^2$ (in this case about 90% of atoms are accumulated in the working levels). Dashed line is for the traditional scheme involving the $\sigma^+$-polarized probe beam on $D_1$-line only.

The results of numerical calculations are shown in Fig. 4. It shows that the amplitude-to-width ratio for the transversal pumping scheme is one order of magnitude higher than for the $\sigma^+$-polarized beam only.

Fig.4(d) shows the amplitude-to-width ratio, divided by square root of intensity (that determines the figure of merit of quantum frequency standard [11] in the case when the noise is the shot noise). Standard stability estimations [11] demonstrate that the best short-term stability attainable with this new scheme is $\sigma_y \leq 10^{-13}/\sqrt{\tau}$.

This research is supported by the INTAS-CNES, project 03-53-5175, and by the Ministry of Education and Science of Russia, project UR.01.01.287.